# DISPLACEMENT AND STRESS ANALYSIS OF AN ELASTIC HOLLOW DISK: COMPARISON WITH STRENGTH OF MATERIALS' PREDICTION

**K. Okamura,[1] Y. Sato,[2] and S. Takada[2]**

[1]Department of Industrial Technology and Innovation, Tokyo University of Agriculture and Technology,
 2-24-16 Naka-cho, Koganei, Tokyo 184-8588, Japan

[2]Department of Mechanical Systems Engineering, Tokyo University of Agriculture and Technology,
 2-24-16 Naka-cho, Koganei, Tokyo 184-8588, Japan

*This paper analyzes the stress distribution in a two-dimensional elastic disk under diametric loading, with a focus on enhancing the understanding of concrete and rock materials' mechanical behavior. The study revisits the Brazilian test and addresses its high shear stress issue near loading points by exploring the ring test, which introduces a central hole in the disk. Using dynamic elasticity theory, we derive stress distributions over time and extend the analysis to static conditions. This approach distinguishes between longitudinal and transverse wave effects, providing a detailed stress field analysis. By drawing parallels with curved beam theories, we demonstrate the applicability of dynamic elasticity theory to complex stress problems, offering improved insights into the stress behavior in elastic disks.*

**Introduction.** Understanding the mechanical strength of concrete and rock materials is crucial in the design and construction of various structures. Among the numerous strength testing methods, the Brazilian test stands out as a widely recognized technique for evaluating the tensile strength of concrete [1, 2]. This method involves applying a concentrated diametric load to a cylindrical specimen, leading to tensile stress perpendicular to the load axis. The uniformity of this tensile stress across the specimen's diameter allows for the determination of failure stress upon the specimen's fracture. However, a significant drawback of the Brazilian test is the high shear stress concentration near the loading points, which can influence the accuracy of the test results [3–5].

To address this issue, the ring test has been proposed, wherein a central hole is introduced to the specimen to mitigate shear stress concentrations [6–8]. The focus of previous studies has been on the stress distribution at the intersection of the hole and the load axis, providing valuable insights into the stress behavior along the load axis. Nevertheless, a comprehensive analysis of the stress distribution throughout the entire interior of the disk has not yet been explicitly presented. It should be noted that the same system we consider here, a two-dimensional elastic hollow disk, has been investigated in a previous study [9, 10], in which the stress distribution has been plotted. However, the details of the theory have not been described, and the validity of the results remains unclear.

Theoretical approaches to these problems commonly employ stress functions, leveraging complex analysis techniques that are well-suited for two-dimensional systems. While powerful, this method is inherently limited to static problems and two-dimensional scenarios [4, 11–14]. In contrast, dynamic elasticity theory introduces potentials to describe stress and displacement fields over time, offering the advantage of distinguishing between longitudinal and transverse wave effects. This approach has been successfully applied to both two-dimensional elastic disks and three-dimensional elastic spheres, elucidating the dynamic stress responses in these geometries [10, 15–19].

A similar analytical framework is employed in the study of curved beams in the field of strength of materials. Curved beams, unlike straight beams, experience non-uniform stress distributions due to their geometry, particularly under bending loads. Stress analysis of curved beams often involves the use of complex functions and stress potentials to accurately describe the stress fields [4, 20–22]. These methods allow for a detailed understanding of the stress concentrations and distributions, which is essential for predicting failure and designing resilient structures. The theoretical principles applied to curved beams can be related to the stress analysis of elastic disks, providing a broader context for understanding the mechanical behavior under various loading conditions.

In this context, this paper revisits the problem of stress analysis in two-dimensional elastic hollow disks within the framework of elastodynamic theory. By examining the long-time behavior of the dynamic solution, we extend the analysis to static stress conditions, thereby providing a more complete understanding of stress distribution in two-dimensional elastic hollow disks under diametric loading. This approach not only enhances the existing knowledge of stress analysis in elastic disks but also bridges the methodologies used for curved beams, demonstrating the versatility and applicability of elastodynamic theory in solving complex stress problems.

**1. Model and Setup.** Let us consider a two-dimensional elastic hollow disk whose outer and inner radii are $R_\text{o}$ and $R_\text{i}$, respectively. The loading starts to act on this disk diametrically at $t = 0$ as shown in Fig. 1. Without loss of generality, we can choose these two points as $\theta = 0$ and $\pi$. Under this condition, we will consider the deformation and stress inside the disk. When we set the magnitude of the loading as $P_0$, the disk should satisfy the following boundary conditions:

$$\sigma_{rr}|_{r=R_\text{o}} = -P(\theta)\Theta(t), \qquad \sigma_{r\theta}|_{r=R_\text{o}} = \sigma_{rr}|_{r=R_\text{i}} = \sigma_{r\theta}|_{r=R_\text{i}} = 0, \tag{1}$$

where $P(\theta)$ can be written as

$$P(\theta) = \frac{P_0}{R_\text{o}}[\delta(\theta) + \delta(\theta - \pi)] = \frac{2P_0}{\pi R_\text{o}}\left[\frac{1}{2} + \sum_{m=2,4,\ldots} \cos(m\theta)\right], \tag{2}$$

with the delta function $\delta(x)$. In the next section, we solve the linearized elastodynamic equation under these boundary conditions.

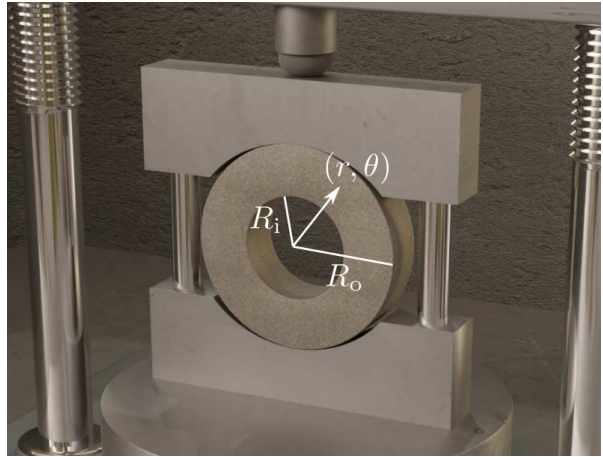

Fig. 1. Schematic of our system. The stress acts on the outer surface of a hollow disk whose outer and inner radii are given by $R_\text{o}$ and $R_\text{i}$, respectively.

**2. Linear Elastodynamics.** The starting point of our analysis is a linear elastodynamic equation. When we assume the plain stress condition, the deformation vector $\boldsymbol{u}$ satisfies the following Navier-Cauchy equations [23]:

$$\varrho_0 \frac{\partial^2}{\partial t^2}\boldsymbol{u} = G\nabla^2 \boldsymbol{u} + G\frac{1+\nu}{1-\nu}\boldsymbol{\nabla}(\boldsymbol{\nabla} \cdot \boldsymbol{u}). \tag{3}$$

Let us introduce the longitudinal and transverse velocities as

$$v_\text{L} \equiv \sqrt{\frac{2}{1-\nu}\frac{G}{\varrho_0}}, \qquad v_\text{T} \equiv \sqrt{\frac{G}{\varrho_0}}, \tag{4}$$

respectively. We also introduce the following dimensionless quantities for later convenience:

$$\boldsymbol{\rho} \equiv \frac{r}{R_{\mathrm{o}}}, \quad \rho_{\mathrm{i}} \equiv \frac{R_{\mathrm{i}}}{R_{\mathrm{o}}}, \quad \tau \equiv \frac{v_{\mathrm{L}} t}{R_{\mathrm{o}}}, \quad \widetilde{\boldsymbol{u}} \equiv \frac{\pi G}{P_0} \boldsymbol{u}, \quad \widetilde{\overleftrightarrow{\sigma}} \equiv \frac{\pi R_{\mathrm{o}}}{P_0} \widetilde{\sigma}, \quad \mu \equiv \frac{v_{\mathrm{L}}}{v_{\mathrm{T}}} = \sqrt{\frac{2}{1-\nu}} (> 1). \tag{5}$$

Then, Eq. (3) is rewritten as

$$\mu^2 \frac{\partial^2}{\partial \tau^2} \widetilde{\boldsymbol{u}} = \nabla_\rho^2 \widetilde{\boldsymbol{u}} + \frac{1+\nu}{1-\nu} \nabla_\rho (\nabla_\rho \cdot \widetilde{\boldsymbol{u}}), \tag{6}$$

with $\nabla_\rho \equiv R_{\mathrm{o}} \nabla$.

From Helmholtz theorem, the (dimensionless) deformation vector can be written in terms of the scalar and vector potentials, $\phi$ and $\boldsymbol{A} = (0,0,A)^{\mathrm{T}}$, respectively, as

$$\widetilde{\boldsymbol{u}} = \nabla_\rho \phi + \nabla_\rho \times \boldsymbol{A}. \tag{7}$$

Substituting Eq. (7) into the Navier-Cauchy equation (6), we find that both potentials should satisfy the wave equations [23]:

$$\nabla_\rho^2 \phi = \frac{\partial^2 \phi}{\partial \tau^2}, \quad \nabla_\rho^2 A = \mu^2 \frac{\partial^2 A}{\partial \tau^2}. \tag{8}$$

To solve Eqs. (8), let us introduce the Laplace transforms of $\phi$ and $A$ as

$$\overline{\phi}(s) \equiv \int_0^\infty \phi \, e^{-s\tau} d\tau, \quad \overline{A}(s) \equiv \int_0^\infty A \, e^{-s\tau} d\tau. \tag{9}$$

When we assume that the system is static for $\tau < 0$, the wave equations (8) become

$$\nabla_\rho^2 \overline{\phi} = s^2 \overline{\phi}, \quad \nabla_\rho^2 \overline{A} = \mu^2 s^2 \overline{A}. \tag{10}$$

These equations are solvable once we adopt the separation of variables. The general solutions are given by

$$\overline{\phi}(s) = \sum_{m=0}^\infty [a_m^I(s) I_m(\rho s) + a_m^K(s) K_m(\rho s)] \cos(m\theta), \tag{11a}$$

$$\overline{A}(s) = \sum_{m=1}^\infty [b_m^I(s) I_m(\mu \rho s) + b_m^K(s) K_m(\mu \rho s)] \sin(m\theta), \tag{11b}$$

with the modified Bessel functions of the first and second kinds, $I_m(\rho)$ and $K_m(\rho)$, respectively [24]:

$$I_m(\rho) = \sum_{n=0}^\infty \frac{1}{n!\, \Gamma(n+m+1)} \left(\frac{\rho}{2}\right)^{2n+m}, \quad K_m(\rho) = \frac{\pi}{2} \frac{I_{-m}(\rho) - I_m(\rho)}{\sin(m\pi)}. \tag{12}$$

Here, the coefficients $a_m^I(s)$, $a_m^K(s)$, $b_m^I(s)$, and $b_m^K(s)$ should be determined to satisfy the boundary condition of the stress given by Eq. (1).

For this purpose, let us first show the expressions of the deformation vector and the stress tensor in terms of the potentials. By adopting the Laplace transforms of $\widetilde{\boldsymbol{u}}$ and $\widetilde{\overleftrightarrow{\sigma}}$ in the same manner with Eq. (9), each component is given by

$$\overline{u}_\rho = \frac{\partial \overline{\phi}}{\partial \rho} + \frac{1}{\rho} \frac{\partial \overline{A}}{\partial \theta}, \quad \overline{u}_\theta = \frac{1}{\rho} \frac{\partial \overline{\phi}}{\partial \theta} - \frac{\partial \overline{A}}{\partial \rho}, \tag{13a}$$

$$\frac{1}{2} \overline{\sigma}_{\rho\rho} = -\frac{1}{\rho} \frac{\partial \overline{\phi}}{\partial \rho} - \frac{1}{\rho^2} \frac{\partial^2 \overline{\phi}}{\partial \theta^2} + \frac{\partial}{\partial \rho}\left(\frac{1}{\rho} \frac{\partial \overline{A}}{\partial \theta}\right) + \frac{1}{1-\nu} \nabla_\rho^2 \overline{\phi}, \tag{13b}$$

$$\frac{1}{2} \overline{\sigma}_{\rho\theta} = -\frac{\partial^2 \overline{A}}{\partial \rho^2} + \frac{\partial}{\partial \rho}\left(\frac{1}{\rho} \frac{\partial \overline{\phi}}{\partial \theta}\right) + \frac{1}{2} \nabla_\rho^2 \overline{A}, \tag{13c}$$

$$\frac{1}{2} \overline{\sigma}_{\theta\theta} = \frac{1}{\rho} \frac{\partial \overline{\phi}}{\partial \rho} + \frac{1}{\rho^2} \frac{\partial^2 \overline{\phi}}{\partial \theta^2} - \frac{\partial}{\partial \rho}\left(\frac{1}{\rho} \frac{\partial \overline{A}}{\partial \theta}\right) + \frac{\nu}{1-\nu} \nabla_\rho^2 \overline{\phi}, \tag{13d}$$

in the polar coordinates. Substituting Eqs. (11) into Eqs. (13), these are rewritten as
$\rho \overline{u}_\rho = -F_{0,0}(\rho s) a_0^I(s) - G_{0,0}(\rho s) a_0^K(s)$

$$+ \sum_{m=1}^{\infty}\left[-F_{m,0}(\rho s)a_m^I(s) + mI_m(\mu\rho s)b_m^I(s) - G_{m,0}(\rho s)a_m^K(s) + mK_m(\mu\rho s)b_m^K(s)\right]\cos(m\theta), \quad (14a)$$

$$\rho\bar{u}_\theta = \sum_{m}^{\infty}\left[-mI_m(\rho s)a_m^I(s) + F_{m,0}(\mu\rho s)b_m^I(s) - mK_m(\rho s)a_m^K(s) + G_{m,0}(\mu\rho s)b_m^K(s)\right]\sin(m\theta), \quad (14b)$$

$$\frac{\rho^2}{2}\bar{\sigma}_{\rho\rho} = F_{0,1}(\rho s)a_0^I(s) + G_{0,1}(\rho s)a_0^K(s)$$

$$+ \sum_{m=1}^{\infty}\left[F_{m,1}(\rho s)a_m^I(s) - F_{m,2}(\mu\rho s)b_m^I(s) + G_{m,1}(\rho s)a_m^K(s) - G_{m,2}(\mu\rho s)b_m^K(s)\right]\cos(m\theta), \quad (14c)$$

$$\frac{\rho^2}{2}\bar{\sigma}_{\rho\theta} = \sum_{m=1}^{\infty}\left[F_{m,2}(\rho s)a_m^I(s) - F_{m,3}(\mu\rho s)b_m^I(s) + G_{m,2}(\rho s)a_m^K(s) - G_{m,3}(\mu\rho s)b_m^K(s)\right]\sin(m\theta), \quad (14d)$$

$$\frac{\rho^2}{2}\bar{\sigma}_{\theta\theta} = F_{0,4}(\rho s)a_0^I(s) + G_{0,4}(\rho s)a_0^K(s)$$

$$+ \sum_{m=1}^{\infty}\left[F_{m,4}(\rho s)a_m^I(s) + F_{m,2}(\mu\rho s)b_m^I(s) + G_{m,4}(\rho s)a_m^K(s) + G_{m,2}(\mu\rho s)b_m^K(s)\right]\cos(m\theta). \quad (14e)$$

Here, we have introduced the following equations:

$$F_{m,0}(z) \equiv -mI_m(z) - zI_{m+1}(z), \quad (15a)$$

$$G_{m,0}(z) \equiv -mK_m(z) + zK_{m+1}(z), \quad (15b)$$

$$F_{m,1}(z) \equiv \left[m(m-1) + \frac{z^2}{1-\nu}\right]I_m(z) - zI_{m+1}(z), \quad (15c)$$

$$G_{m,1}(z) \equiv \left[m(m-1) + \frac{z^2}{1-\nu}\right]K_m(z) + zK_{m+1}(z), \quad (15d)$$

$$F_{m,2}(z) \equiv -m(m-1)I_m(z) - mzI_{m+1}(z), \quad (15e)$$

$$G_{m,2}(z) \equiv -m(m-1)K_m(z) + mzK_{m+1}(z), \quad (15f)$$

$$F_{m,3}(z) \equiv \left[m(m-1) + \frac{z^2}{2}\right]I_m(z) - zI_{m+1}(z), \quad (15g)$$

$$G_{m,3}(z) \equiv \left[m(m-1) + \frac{z^2}{2}\right]K_m(z) + zK_{m+1}(z), \quad (15h)$$

$$F_{m,4}(z) \equiv \left[-m(m-1) + \frac{\nu z^2}{1-\nu}\right]I_m(z) + zI_{m+1}(z), \quad (15i)$$

$$G_{m,4}(z) \equiv \left[-m(m-1) + \frac{\nu z^2}{1-\nu}\right]K_m(z) - zK_{m+1}(z). \quad (15j)$$

Equations (14c) and (14d) should satisfy the boundary conditions (1) in the Laplace space:

$$\frac{1}{2}\bar{\sigma}_{\rho\rho}|_{\rho=1} = -\left[\frac{1}{2s} + \frac{1}{s}\sum_{m=2,4,\cdots}\cos(m\theta)\right], \quad (16a)$$

$$\bar{\sigma}_{\rho\theta}|_{\rho=1} = \bar{\sigma}_{\rho\rho}|_{\rho=\rho_i} = \bar{\sigma}_{\rho\theta}|_{\rho=\rho_i} = 0. \quad (16b)$$

We can first find that

$$a_m^I(s) = b_m^I(s) = a_m^K(s) = b_m^K(s) = 0, \quad (17)$$

for $m = 1, 3, 5, \cdots$. Next, let us consider the case for $m = 0$. The coefficients $a_0^I(s)$ and $a_0^K(s)$ should satisfy

$$F_{0,1}(s)a_0^I(s) + G_{0,1}(s)a_0^K(s) = -\frac{1}{2s}, \quad (18a)$$

$$F_{0,1}(\rho_i s)a_0^I(s) + G_{0,1}(\rho_i s)a_0^K(s) = 0. \quad (18b)$$

Then, we get

$$a_0^I(s) = -\frac{1}{2s}\frac{G_{0,1}(\rho_i s)}{F_{0,1}(s)G_{0,1}(\rho_i s) - F_{0,1}(\rho_i s)G_{0,1}(s)}, \tag{19a}$$

$$a_0^K(s) = \frac{1}{2s}\frac{F_{0,1}(\rho_i s)}{F_{0,1}(s)G_{0,1}(\rho_i s) - F_{0,1}(\rho_i s)G_{0,1}(s)}. \tag{19b}$$

Third, for $m = 2,4,\cdots$, the four coefficients should satisfy

$$F_{m,1}(s)a_m^I(s) - F_{m,2}(\mu s)b_m^I(s) + G_{m,1}(s)a_m^K(s) - G_{m,2}(\mu s)b_m^K(s) = -\frac{1}{s}, \tag{20a}$$

$$F_{m,1}(\rho_i s)a_m^I(s) - F_{m,2}(\mu\rho_i s)b_m^I(s) + G_{m,1}(\rho_i s)a_m^K(s) - G_{m,2}(\mu\rho_i s)b_m^K(s) = 0, \tag{20b}$$

$$F_{m,2}(s)a_m^I(s) - F_{m,3}(\mu s)b_m^I(s) + G_{m,2}(s)a_m^K(s) - G_{m,3}(\mu s)b_m^K(s) = 0, \tag{20c}$$

$$F_{m,2}(\rho_i s)a_m^I(s) - F_{m,3}(\mu\rho_i s)b_m^I(s) + G_{m,2}(\rho_i s)a_m^K(s) - G_{m,3}(\mu\rho_i s)b_m^K(s) = 0. \tag{20d}$$

Then, the coefficients can be determined as

$$a_m^I(s) = -\frac{N_{m,1}(s)}{sD_m(s)}, \quad b_m^I(s) = -\frac{N_{m,2}(s)}{sD_m(s)}, \quad a_m^K(s) = -\frac{N_{m,3}(s)}{sD_m(s)}, \quad b_m^K(s) = -\frac{N_{m,4}(s)}{sD_m(s)}, \tag{21}$$

where $D_m(s)$ and $N_{m,i}(s)$ ($i = 1,2,3,4$) are given by the following determinants of the matrices:

$$D_m(s) = \begin{vmatrix} F_{m,1}(s) & F_{m,2}(\mu s) & G_{m,1}(s) & G_{m,2}(\mu s) \\ F_{m,1}(\rho_i s) & F_{m,2}(\mu\rho_i s) & G_{m,1}(\rho_i s) & G_{m,2}(\mu\rho_i s) \\ F_{m,2}(s) & F_{m,3}(\mu s) & G_{m,2}(s) & G_{m,3}(\mu s) \\ F_{m,2}(\rho_i s) & F_{m,3}(\mu\rho_i s) & G_{m,2}(\rho_i s) & G_{m,3}(\mu\rho_i s) \end{vmatrix}, \tag{22a}$$

$$N_{m,1}(s) = \begin{vmatrix} F_{m,2}(\mu\rho_i s) & G_{m,1}(\rho_i s) & G_{m,2}(\mu\rho_i s) \\ F_{m,3}(\mu s) & G_{m,2}(s) & G_{m,3}(\mu s) \\ F_{m,3}(\mu\rho_i s) & G_{m,2}(\rho_i s) & G_{m,3}(\mu\rho_i s) \end{vmatrix}, \tag{22b}$$

$$N_{m,2}(s) = \begin{vmatrix} F_{m,1}(\rho_i s) & G_{m,1}(\rho_i s) & G_{m,2}(\mu\rho_i s) \\ F_{m,2}(s) & G_{m,2}(s) & G_{m,3}(\mu s) \\ F_{m,2}(\rho_i s) & G_{m,2}(\rho_i s) & G_{m,3}(\mu\rho_i s) \end{vmatrix}, \tag{22c}$$

$$N_{m,3}(s) = \begin{vmatrix} F_{m,1}(\rho_i s) & F_{m,2}(\mu\rho_i s) & G_{m,2}(\mu\rho_i s) \\ F_{m,2}(s) & F_{m,3}(\mu s) & G_{m,3}(\mu s) \\ F_{m,2}(\rho_i s) & F_{m,3}(\mu\rho_i s) & G_{m,3}(\mu\rho_i s) \end{vmatrix}, \tag{22d}$$

$$N_{m,4}(s) = \begin{vmatrix} F_{m,1}(\rho_i s) & F_{m,2}(\mu\rho_i s) & G_{m,1}(\rho_i s) \\ F_{m,2}(s) & F_{m,3}(\mu s) & G_{m,2}(s) \\ F_{m,2}(\rho_i s) & F_{m,3}(\mu\rho_i s) & G_{m,2}(\rho_i s) \end{vmatrix}. \tag{22e}$$

Substituting Eqs. (22) into Eqs. (14) and performing the inverse Laplace transforms, each component of the (dimensionless) deformation vector and the (dimensionless) stress tensor is given by

$$\begin{Bmatrix} \tilde{u}_\rho \\ \tilde{\sigma}_{\rho\rho} \\ \tilde{\sigma}_{\theta\theta} \end{Bmatrix} = \sum_{m=0,2,4,\cdots}^{\infty} \begin{Bmatrix} \tilde{u}_\rho^{(m)} \\ \tilde{\sigma}_{\rho\rho}^{(m)} \\ \tilde{\sigma}_{\theta\theta}^{(m)} \end{Bmatrix} \cos(m\theta), \quad \begin{Bmatrix} \tilde{u}_\theta \\ \tilde{\sigma}_{\rho\theta} \end{Bmatrix} = \sum_{m=2,4,\cdots}^{\infty} \begin{Bmatrix} \tilde{u}_\theta^{(m)} \\ \tilde{\sigma}_{\rho\theta}^{(m)} \end{Bmatrix} \sin(m\theta), \tag{23}$$

with

$$\tilde{u}_\rho^{(0)} = \frac{1}{2\rho}\frac{1}{2\pi i}\int_{\text{Br}}\frac{1}{s}\frac{F_{0,0}(\rho s)G_{0,1}(\rho_i s) - F_{0,1}(\rho_i s)G_{0,0}(\rho s)}{F_{0,1}(s)G_{0,1}(\rho_i s) - F_{0,1}(\rho_i s)G_{0,1}(s)}e^{s\tau}ds, \tag{24a}$$

$$\tilde{\sigma}_{\rho\rho}^{(0)} = -\frac{1}{\rho^2}\frac{1}{2\pi i}\int_{\text{Br}}\frac{1}{s}\frac{F_{0,1}(\rho s)G_{0,1}(\rho_i s) - F_{0,1}(\rho_i s)G_{0,1}(\rho s)}{F_{0,1}(s)G_{0,1}(\rho_i s) - F_{0,1}(\rho_i s)G_{0,1}(s)}e^{s\tau}ds, \tag{24b}$$

$$\tilde{\sigma}_{\theta\theta}^{(0)} = -\frac{1}{\rho^2}\frac{1}{2\pi i}\int_{\text{Br}}\frac{1}{s}\frac{F_{0,4}(\rho s)G_{0,1}(\rho_i s) - F_{0,1}(\rho_i s)G_{0,4}(\rho s)}{F_{0,1}(s)G_{0,1}(\rho_i s) - F_{0,1}(\rho_i s)G_{0,1}(s)}e^{s\tau}ds, \tag{24c}$$

and

$$\tilde{u}_\rho^{(m)} = \frac{1}{\rho} \frac{1}{2\pi i} \int_{\text{Br}} \frac{F_{m,0}(\rho s)N_{m,1}(s) - mI_m(\mu\rho s)N_{m,2}(s) + G_{m,0}(\rho s)N_{m,3}(s) - mK_m(\mu\rho s)N_{m,4}(s)}{sD_m(s)} e^{s\tau} ds, \quad (25a)$$

$$\tilde{u}_\theta^{(m)} = \frac{1}{\rho} \frac{1}{2\pi i} \int_{\text{Br}} \frac{mI_m(\rho s)N_{m,1}(s) - F_{m,0}(\mu\rho s)N_{m,2}(s) + mK_m(\rho s)N_{m,3}(s) - G_{m,0}(\mu\rho s)N_{m,4}(s)}{sD_m(s)} e^{s\tau} ds, \quad (25b)$$

$$\tilde{\sigma}_{\rho\rho}^{(m)} = -\frac{2}{\rho^2} \frac{1}{2\pi i} \int_{\text{Br}} \frac{F_{m,1}(\rho s)N_{m,1}(s) - F_{m,2}(\mu\rho s)N_{m,2}(s) + G_{m,1}(\rho s)N_{m,3}(s) - G_{m,2}(\mu\rho s)N_{m,4}(s)}{sD_m(s)} e^{s\tau} ds, \quad (25c)$$

$$\tilde{\sigma}_{\rho\theta}^{(m)} = -\frac{2}{\rho^2} \frac{1}{2\pi i} \int_{\text{Br}} \frac{F_{m,2}(\rho s)N_{m,1}(s) - F_{m,3}(\mu\rho s)N_{m,2}(s) + G_{m,2}(\rho s)N_{m,3}(s) - G_{m,3}(\mu\rho s)N_{m,4}(s)}{sD_m(s)} e^{s\tau} ds, \quad (25d)$$

$$\tilde{\sigma}_{\theta\theta}^{(m)} = -\frac{2}{\rho^2} \frac{1}{2\pi i} \int_{\text{Br}} \frac{F_{m,4}(\rho s)N_{m,1}(s) + F_{m,2}(\mu\rho s)N_{m,2}(s) + G_{m,4}(\rho s)N_{m,3}(s) + G_{m,2}(\mu\rho s)N_{m,4}(s)}{sD_m(s)} e^{s\tau} ds, \quad (25e)$$

for $m = 2, 4, \cdots$. It is noted that $\int_{\text{Br}} \cdots = \int_{\gamma-i\infty}^{\gamma+i\infty} \cdots$ is the Bromwich integral to calculate the inverse Laplace transform. Here, $\gamma(>0)$ should be larger than any real parts of the pole of the integrands in Eqs. (24) and (25).

**3. Static Solutions.** In this section, let us obtain the static solutions. These static solutions are realized in the long-time limit $\tau \to \infty$, which can be evaluated in terms of the final value theorem of the Laplace transforms, $\lim_{\tau \to \infty} f(\tau) = \lim_{s \to 0}[s\overline{f}(s)]$. After some calculations with the aid of l'Hopital's rule, we get

$$\lim_{\tau \to \infty} \tilde{u}_\rho^{(0)} = \frac{1}{2\rho} \lim_{\tau \to \infty} \frac{F_{0,0}(\rho s)G_{0,1}(\rho_i s) - F_{0,1}(\rho_i s)G_{0,0}(\rho s)}{F_{0,1}(s)G_{0,1}(\rho_i s) - F_{0,1}(\rho_i s)G_{0,1}(s)} = -\frac{(1-\nu)\rho^2 + (1+\nu)\rho_i^2}{2(1+\nu)\rho(1-\rho_i^2)}, \quad (26a)$$

$$\lim_{\tau \to \infty} \tilde{\sigma}_{\rho\rho}^{(0)} = -\frac{1}{\rho^2} \lim_{\tau \to \infty} \frac{F_{0,1}(\rho s)G_{0,1}(\rho_i s) - F_{0,1}(\rho_i s)G_{0,1}(\rho s)}{F_{0,1}(s)G_{0,1}(\rho_i s) - F_{0,1}(\rho_i s)G_{0,1}(s)} = -\frac{\rho^2 - \rho_i^2}{\rho^2(1-\rho_i^2)}, \quad (26b)$$

$$\lim_{\tau \to \infty} \tilde{\sigma}_{\theta\theta}^{(0)} = -\frac{1}{\rho^2} \lim_{\tau \to \infty} \frac{F_{0,4}(\rho s)G_{0,1}(\rho_i s) - F_{0,1}(\rho_i s)G_{0,4}(\rho s)}{F_{0,1}(s)G_{0,1}(\rho_i s) - F_{0,1}(\rho_i s)G_{0,1}(s)} = -\frac{\rho^2 + \rho_i^2}{\rho^2(1-\rho_i^2)}, \quad (26c)$$

for $m = 0$, and

$$\lim_{\tau \to \infty} \tilde{u}_\rho^{(m)} = \frac{\mathcal{N}_m^{(1)}}{2(1+\nu)(m^2-1)\rho \mathcal{D}_m}, \quad \lim_{\tau \to \infty} \tilde{u}_\theta^{(m)} = \frac{\mathcal{N}_m^{(2)}}{2(1+\nu)(m^2-1)\rho \mathcal{D}_m}, \quad (27a)$$

$$\lim_{\tau \to \infty} \tilde{\sigma}_{\rho\rho}^{(m)} = \frac{\mathcal{N}_m^{(3)}}{\rho^2 \mathcal{D}_m}, \quad \lim_{\tau \to \infty} \tilde{\sigma}_{\rho\theta}^{(m)} = \frac{\mathcal{N}_m^{(4)}}{\rho^2 \mathcal{D}_m}, \quad \lim_{\tau \to \infty} \tilde{\sigma}_{\theta\theta}^{(m)} = \frac{\mathcal{N}_m^{(5)}}{\rho^2 \mathcal{D}_m}, \quad (27b)$$

for $m = 2, 4, \cdots$, where $\mathcal{D}_m$ and $\mathcal{N}_m^{(i)}$ ($i=1, 2, \ldots, 5$) are given by

$$\mathcal{D}_m \equiv m^2 \rho_i^{2m-2}(1-\rho_i^2)^2 - (1-\rho_i^{2m})^2, \quad (28a)$$

$$\mathcal{N}_m^{(1)} \equiv -m(m-1)(1+\nu)\mathcal{C}_m^{(1)} + m(m+1)(1+\nu)\mathcal{C}_m^{(2)} - (m-1)[m(1+\nu) - 2(1-\nu)]\mathcal{C}_m^{(3)}$$
$$+ (m+1)[m(1+\nu) + 2(1-\nu)]\mathcal{C}_m^{(4)}, \quad (28b)$$

$$\mathcal{N}_m^{(2)} \equiv -m(m-1)(1+\nu)\mathcal{C}_m^{(1)} - m(m+1)(1+\nu)\mathcal{C}_m^{(2)} + (m-1)[m(1+\nu) - 2(1-\nu)]\mathcal{C}_m^{(3)}$$
$$+ (m+1)[m(1+\nu) + 2(1-\nu)]\mathcal{C}_m^{(4)}, \quad (28c)$$

$$\mathcal{N}_m^{(3)} \equiv m\mathcal{C}_m^{(1)} + m\mathcal{C}_m^{(2)} - (m-2)\mathcal{C}_m^{(3)} - (m+2)\mathcal{C}_m^{(4)}, \quad (28d)$$

$$\mathcal{N}_m^{(4)} \equiv \mathcal{C}_m^{(1)} - \mathcal{C}_m^{(2)} + \mathcal{C}_m^{(3)} - \mathcal{C}_m^{(4)} \quad (28e)$$

$$\mathcal{N}_m^{(5)} \equiv m\mathcal{C}_m^{(1)} + m\mathcal{C}_m^{(2)} - (m+2)\mathcal{C}_m^{(3)} - (m-2)\mathcal{C}_m^{(4)}, \quad (28f)$$

with

$$\mathcal{C}_m^{(1)} \equiv \rho_i^m \left(\frac{\rho_i}{\rho}\right)^m [m(1-\rho_i^2) + 1 - \rho_i^{2m}], \quad \mathcal{C}_m^{(2)} \equiv \rho^m[1 - \rho_i^{2m} + m\rho_i^{2m}(1-\rho_i^2)], \quad (29a)$$

$$\mathcal{C}_m^{(3)} \equiv \rho^{m+2}[1 - \rho_i^{2m} + m\rho_i^{2m-2}(1-\rho_i^2)], \quad \mathcal{C}_m^{(4)} \equiv \rho_i^m \left(\frac{\rho_i}{\rho}\right)^{m-2} [m(1-\rho_i^2) + \rho_i^2(1-\rho_i^{2m})]. \quad (29b)$$

It should be noted that these solutions are consistent with the previous studies for solid disks [4, 25, 26] when we consider $\rho_i \to 0$. In the following, we simply put $\lim_{\tau \to \infty} \tilde{u}_\alpha$ and $\lim_{\tau \to \infty} \tilde{\sigma}_{\alpha\beta}$ as $\tilde{u}_\alpha$ and $\tilde{\sigma}_{\alpha\beta}$, respectively, because we are only interested in the static quantities in this paper.

Let us introduce the magnitude of the dimensionless deformation $\tilde{u}$ and the second stress difference $\Delta\tilde{\sigma}$, where they are defined as

$$\tilde{u} \equiv \sqrt{\tilde{u}_\rho^2 + \tilde{u}_\theta^2}, \qquad \Delta\tilde{\sigma} \equiv \sqrt{(\tilde{\sigma}_{\rho\rho} - \tilde{\sigma}_{\theta\theta})^2 + 4\tilde{\sigma}_{\rho\theta}^2}, \tag{30}$$

respectively.

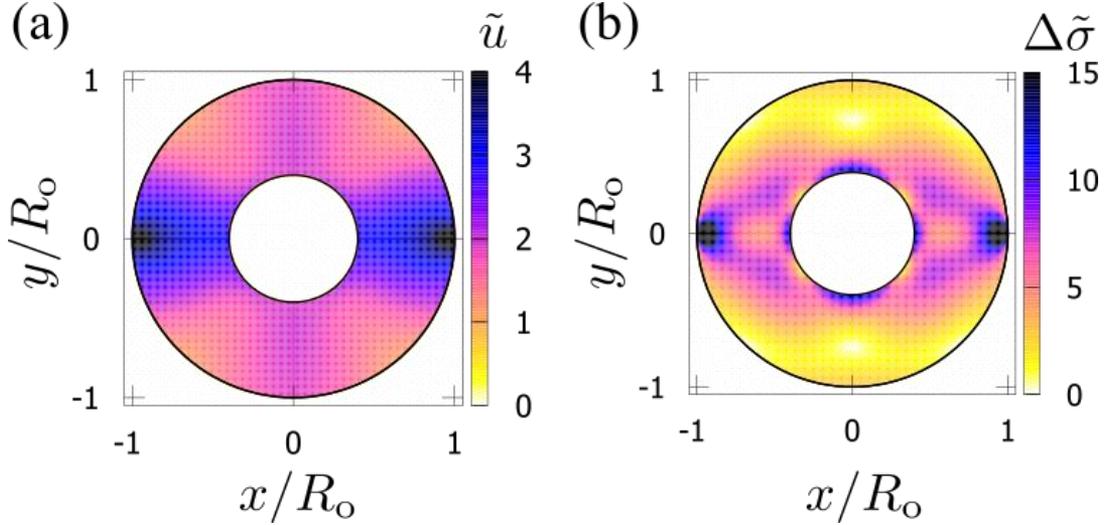

Fig. 2. Plots of the (dimensionless) (a) displacement $\tilde{u}$ and (b) principal stress difference $\Delta\tilde{\sigma}$ evaluated from Eq. (30) for $\nu = 0.3$ and $\rho_i = 0.4$. The color indicates the magnitude of each quantity.

Figure 2 shows the results of the (dimensionless) displacement $\tilde{u}$ and principal stress difference $\Delta\tilde{\sigma}$ evaluated from Eq. (30) for $\nu = 0.3$ and $\rho_i = 0.4$. From Figure 2(a), the displacement exhibits minima in the $\theta = \pm\pi/4$ directions. This behavior is almost identical to that of a solid disk [4]. On the other hand, focusing on the stress in Figure 2(b), there are maxima near the inner edge in the direction of the loading line and perpendicular to it. Additionally, on the loading line, there are minima near the loading points and the inner edge. This is a complex structure not observed in solid disks [4], which appears due to the presence of the inner edge.

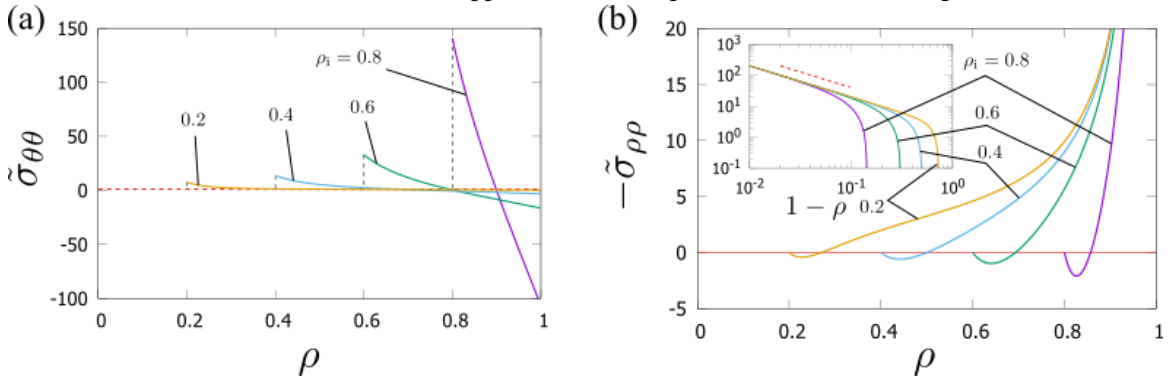

Fig. 3. Plots of the (dimensionless) (a) tensile and (b) compressive stresses along the loading line for various $\rho_i$ with $\nu = 0.3$. The inset of panel (b) is the plot against $1 - \rho$. The dashed line represents $(1 - \rho)^{-1}$.

In relation to Figure 2, we focus on the stress on the loading line, which is of particular interest in strength tests. On the loading line, the tensile stress perpendicular to this line, and the compressive stress parallel to the loading line are illustrated in Fig. 3(a) and (b), respectively. It is known that in solid disks, the tensile stress is constant regardless of position, which forms the theoretical basis for the Brazilian test conducted on concrete. In contrast, in hollow disks, this stress is not constant, and a switch from tensile to compressive stress is observed at certain points (Fig. 3 (a)). This behavior has also been reported experimentally. This is because, being hollow, the region near the inner edge can deform inward. As a result, the outer edge region deforms by being dragged by the load. This is the origin of the compression near the outer edge. Meanwhile, near the inner edge, the force is applied in a manner that pushes it out from the outer edge. However, due to the need to satisfy the free boundary condition (1), the force needs to be released in the circumferential direction. This leads to the generation of tensile stress. A similar discussion can be applied to understand the tensile region near the inner edge parallel to the load axis (Fig. 3(b)). Here, we further consider the relationship between stress and displacement. Considering the stress-strain relationship and the strain-displacement relationship, the normal stress in the radial direction $\tilde{\sigma}_{\rho\rho}$ is given by

$$\tilde{\sigma}_{\rho\rho} = \frac{2G}{1-2\nu}\left[(1-\nu)\frac{\partial \tilde{u}_\rho}{\partial \rho} + \nu\left(\frac{1}{\rho}\frac{\partial \tilde{u}_\theta}{\partial \theta} + \frac{\tilde{u}_\rho}{\rho}\right)\right]. \tag{31}$$

In polar coordinates, unlike a rectangular element, the increase in $\tilde{u}_\theta$ in the circumferential direction affects the small elements. It can be seen that the tensile stress appears due to the influence of this increase.

Furthermore, a divergent behavior caused by the concentrated load is observed near the outer edge on the load axis ($\theta = 0$). This behavior follows $(1-\rho)^{-1}$ as shown in the inset of Fig. 3(b). This can be understood from Eq. (23) as

$$\tilde{\sigma}_{\rho\rho}(\rho, \theta = 0) = \sum_{m=0,2,4,\cdots} \tilde{\sigma}_{\rho\rho}^{(m)} \approx \sum_{m=0,2,4,\cdots} \rho^m = \frac{1}{1-\rho^2} \sim \frac{1}{1-\rho}, \tag{32}$$

which is consistent with the results.

**4. Comparison with Curved Beam in Strength of Materials.** Because an elastic hollow disk can be regarded as a curved beam, we will compare our results with those known for curved beams in the field of the strength of materials. Let us consider a circular ring subjected to a diametric loading. The geometry is taken as in the systems of the previous sections of this paper. According to the strength of materials [20], the maximum deformations in the direction parallel and perpendicular to the loading occur at $\theta = 0$ and $\pi$, respectively, which are given by

$$u_{\rho,\max}^{\parallel} = -\frac{P_0 R_o}{AE\kappa}\left[\frac{\pi}{8} - \frac{1}{\pi(\kappa+1)}\right], \tag{33a}$$

$$u_{\rho,\max}^{\perp} = \frac{P_0 R_o}{AE\kappa}\left[\frac{1}{\pi(\kappa+1)} - \frac{1}{4}\right], \tag{33b}$$

where $A$ is the cross section of the beam, $E = 2(1+\nu)G$ is Young's modulus, and $\kappa$ is a coefficient related to the section, which is defined by [20]

$$\kappa \equiv \frac{1}{1-\rho_i}\log\frac{3-\rho_i}{1+\rho_i} - 1 = \sum_n^\infty \frac{1}{2n+1}\left(\frac{1-\rho_i}{2}\right)^{2n}. \tag{34}$$

When the beam is thin enough as compared to the curvature of the beam $R_o$, that is $1 - \rho_i \ll 1$, the beam is regarded as a thin curved beam. In this case, $\kappa$ is much smaller than unity, and the results become simpler. We can put $1 + \kappa \simeq 1$ and $\kappa \simeq (1-\rho_i)^2/12$. Then, Eqs. (33), respectively, become

$$u_{\rho,\max}^{\parallel} \simeq u_{\rho,\max}^{\parallel(\text{thin})} \equiv -\frac{6\pi}{(1+\nu)(1-\rho_i)^3}\left(\frac{\pi}{8} - \frac{1}{\pi}\right), \tag{35a}$$

$$u_{\rho,\max}^{\perp} \simeq u_{\rho,\max}^{\perp(\text{thin})} \equiv \frac{6\pi}{(1+\nu)(1-\rho_i)^3}\left(\frac{1}{\pi} - \frac{1}{4}\right). \tag{35b}$$

The results of a thin curved beam (35) from the strength of materials can be derived from our solution (23) with Eqs. (26) and (27). Now, let us consider $\tilde{u}_\rho(\rho)$ at the outer edge ($\rho = 1$) of the disk. Under the condition

$$\delta \equiv 1 - \rho_i \ll 1, \tag{36}$$

$\tilde{u}_\rho^{(0)}(1)$ and $\tilde{u}_\rho^{(m)}(1)$ ($m = 2, 4, \cdots$), respectively, become

$$\tilde{u}_\rho^{(0)}(1) = -\frac{(1-\nu)+(1+\nu)\rho_i^2}{2(1+\nu)(1-\rho_i^2)} = \mathcal{O}(\delta^{-1}), \tag{37a}$$

$$\tilde{u}_\rho^{(m)}(1) = -\frac{12}{(1+\nu)(m^2-1)^2\delta^3} + \frac{18}{(1+\nu)(m^2-1)^2\delta^2} + \cdots$$
$$= -\frac{12}{(1+\nu)(m^2-1)^2\delta^3}\left(1-\frac{3}{2}\delta\right) + \mathcal{O}(\delta^{-1}). \tag{37b}$$

Then, after summing up these terms, we get

$$\tilde{u}_\rho(1,\theta) = \sum_{m=0,2,4,\cdots} \tilde{u}_\rho^{(m)}(1)\cos(m\theta) = -\frac{12}{(1+\nu)\delta^3}\left(1-\frac{3}{2}\delta\right)\sum_{m=2,4,\cdots}\frac{1}{(m^2-1)^2}\cos(m\theta) + \mathcal{O}(\delta^{-1})$$

$$\to \begin{cases} u_{\rho,\max}^{\|(\text{thin})}\left(1-\frac{3}{2}\delta\right) & (\theta=0) \\ u_{\rho,\max}^{\perp(\text{thin})}\left(1-\frac{3}{2}\delta\right) & \left(\theta=\frac{\pi}{2}\right) \end{cases} \tag{38}$$

To derive the above, we have used the following identity for any real function $f(m)$:

$$\sum_m f(m)\cos(m\theta) = \Re\left[\sum_m f(m)e^{im\theta}\right]. \tag{39}$$

This means that the leading term corresponds to the results of a thin curved beam from the strength of materials, and the next lowest order is the correction from the leading term.

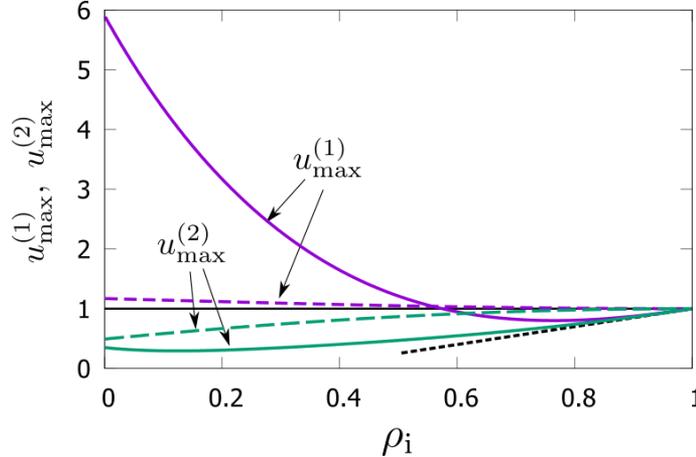

Fig. 4. Plots of $\tilde{u}_{\max}^{(1)}$ and $\tilde{u}_{\max}^{(2)}$ against $\rho_i$ for $\nu = 0.3$. The solid lines represent the ratios of our theoretical results to the results of a thin curved beam from the strength of materials (35). The dashed lines are the ratios of the results of a curved beam with finite width (33) to those of a thin curved beam (35). The dotted line is the first-order correction (38).

Let us introduce the ratios of our theoretical results to the results of a thin curved beam from the strength of materials (35) as

$$\tilde{u}_{\max}^{(1)} \equiv \frac{\tilde{u}_\rho}{u_{\rho,\max}^{\|(\text{thin})}}, \qquad \tilde{u}_{\max}^{(2)} \equiv \frac{\tilde{u}_\rho}{u_{\rho,\max}^{\perp(\text{thin})}}. \tag{40}$$

Figure 4 clearly shows that the results of the strength of materials only applicable near $1-\rho_i \ll 1$. It is also interesting that the results of a curved beam with finite width (33) is almost same as those of a thin curved width and the discrepancy from our theory is large even near $1-\rho_i \ll 1$. In addition, this figure suggests that the first order correction (38) works well for $\delta \lesssim 0.2$.

**Conclusion and Discussion.** In this paper, we have successfully derived the displacement and stress of a two-dimensional elastic hollow disk under a diametric loading from the framework of the elastodynamic theory. Our results clearly have shown the places where the displacement and stress maxima and minima appear in the hollow disk. The signs of tensile and compressive stresses and their dominant locations are also clarified by focusing on the loading line. In addition, our results suggest that the traditional treatment using the strength of materials fails to include a finite width effect to curved beams.

Although we have only considered the static problem, our treatment can be expanded to the dynamic response. If we properly treat the inverse Laplace transforms of the deformation and stress, we can derive the time evolution of them. These calculations seem cumbersome, but this should be our next task.

**Acknowledgment.** The authors thank Shintaro Hokada for his comments and helps on visualization. This work is partially supported by the Grant-in-Aid of MEXT for Scientific Research (Grant No. 24K06974 and No. 24K07193).

**Conflict of Interests.** All authors declare that they have no conflicts of interest.

**Author Contributions.** All authors contributed to the study conception and design. The preliminary calculations were performed by Yosuke Sato. The main calculations and analysis were performed by Ken Okamura and Satoshi Takada. The first draft of the manuscript was written by Satoshi Takada and all authors commented on previous versions of the manuscript. All authors read and approved the final manuscript.

## REFERENCES

1. D. W. Hobbs, "The tensile strength of rocks" Int. J. Rock. Mech. Min. Sci. Geomech. Abstr. Vol. 1, 385–388 (1964).
2. J. Claesson and B. Bohloli, "Brazilian test: stress field and tensile strength of anisotropic rocks using an analytical solution" Int. J. Rock Mech. Min. Sci., Vol. 39, 8 (2002).
3. J. C. Jaeger and N. G. W. Cook, *Fundamentals of Rock Mechanics*, Chapman and Hall (1979).
4. S. P. Timoshenko and J. N. Goodier, *Theory of Elasticity 3rd ed.*, McGraw-Hill (1970).
5. K. Ramesh and K. Shins, "Stress field equations for a disk subjected to self-equilibrated arbitrary loads: revisited" Granul. Matter, Vol. 24, 49 (2022).
6. G. Hondros, "The evaluation of Poisson's ratio and the modulus of materials of a low tensile resistance by the Brazilian (indirect tensile) test with particular reference to concrete" Aust. J. Appl. Sci., Vol. 10, 243–268 (1959).
7. H. Laurent, R. Grèze, P. Y. Manach, S. Thuillier, "Influence of constitutive model in springback prediction using the split-ring test" Int. J. Mech. Sci., Vol. 51, 233–245 (2009).
8. Y. Gao, J. Zhang, P. Han, "Determination of stress relaxation parameters of concrete in tension at early-age by ring test" Constr. Build. Mater., Vol. 41, 152–164 (2013).
9. Y. Hiramatsu and Y. Oka, "Disc Test, Ring Test, Rectangular Plate Test and Irregular Specimen Test for Determining the Tensile Strentgh of Rocks" in *Proceedings of 2nd Congress of the International Society for Rock Mechanics, Belgrade*, Vol.~2, pp.~199−206 (1970).
10. S. Ujihashi and H. Matsumoto, "Dynamic Stresses and Deformations in a Thick Cylinder under Nonaxially Symmetrical Impulsive Loads: 1st Report, Analysis Based on the Two Dimensional Dynamic Theory of Elasticity" Bulletin JSME, Vol. 17, 1418–1425 (1974).
11. N. I. Muskhelishvili, *Some Basic Problems of the Mathematical Theory of Elasticity*, Noordhoff (1953).
12. S. Y. Wang, S. W. Sloan, and C. A. Tang, "Three-Dimensional Numerical Investigations of the Failure Mechanism of a Rock Disc with a Central or Eccentric Hole" Rock Mech. Rock Eng., Vol. 47, 2117 (1985).
13. K. Schonert, "Breakage of spheres and circular discs" Powder Technol., Vol. 143, 2–18 (2004).
14. S. Z. Wu and K. T. Chau, "Dynamic response of an elastic sphere under diametral impacts" Mech. Mater., 38, 1039 (2006).

15. A. E. H. Love, *A Treatise on the Mathematical Theory of Elasticity*, Dover (1944).
16. A. C. Eringen and E. S. Şuhubi, *Elastodynamics, Vol. II Linear Theory*, Academic Press (1975).
17. T. Hua and R. A. Van Gorder, "Wave propagation and pattern formation in two-dimensional hexagonally-packed granular crystals under various configurations" Granul. Matter, Vol. 21, 3 (2019).
18. T. Jingu, K. Hisada, I. Nakahara, and S. Machida, "Transient Stress in a Circular Disk under Diametrical Impact Loads" Bulletin JSME, Vol. 28, 13–23 (1985).
19. D. Kessler and D. Kosloff, "Elastic wave propagation using cylindrical coordinates" Geophys., Vol. 56, 2080–2089 (1991).
20. S. P. Timoshenko and D. H. Young, *Elements of Strength of Materials 4th ed.*, D. Van Nostrand Company (1962).
21. J. A. Ewing, *The strength of materials*, Cambridge University Press (1903).
22. L. D. Landau, L. P. Pitaevkii, A. M. Kosevichm, and E. M. Lifshitz, *Theory of Elasticity*, Elsevier (1986).
23. Y. C. Fung and P. Tong, *Classical and Computational Solid Mechanics*, World Scientific (2001).
24. M. Abramowitz and I. A. Stegun, *Handbook of Mathematical Functions: With Formulas, Graphs, and Mathematical Tables*, Dover (1964).
25. Y. Sato and S. Takada, "Revisiting stress propagation in a three-dimensional elastic sphere under diametric loading" [in Japanese], Trans. JSME, Vol. 90, 23-00262 (2024).
26. Y. Sato, H. Ishikawa, and S. Takada, "Revisiting stress propagation in a two-dimensional elastic circular disk under diametric loading" J. Elast., Vol. 156, 193 (2024).